\documentclass[dvips]{article}
\usepackage{graphicx}
\usepackage{amssymb}
\usepackage{graphicx}
\usepackage{dcolumn}
\usepackage{amsmath}
\usepackage{lscape}
\usepackage{longtable}
\usepackage{subfigure}
\usepackage{color}

\begin{document}
\newcommand{\bd}{\begin{document}}
\newcommand{\ed}{\end{document}}
\newcommand{\bc}{\begin{center}}
\newcommand{\ec}{\end{center}}
\newcommand{\bfr}{\begin{flushright}}
\newcommand{\efr}{\end{flushright}}
\newcommand{\lt}{\left}
\newcommand{\rt}{\right}
\newcommand{\vs}{\vspace}
\newcommand{\hs}{\hspace}
\newcommand{\beq}{\begin{equation}}
\newcommand{\eeq}{\end{equation}}
\newcommand{\lb}{\linebreak}
\newcommand{\pb}{\pagebreak}
\newcommand{\mb}{\makebox}
\newcommand{\fb}{\framebox}
\newcommand{\mc}{\multicolumn}
\newcommand{\ben}{\begin{enumerate}}
\newcommand{\een}{\end{enumerate}}
\newcommand{\bit}{\begin{itemize}}
\newcommand{\eit}{\end{itemize}}
\newcommand{\ol}{\overline}
\newcommand{\un}{\underline}
\newcommand{\lefq}{\lefteqn}
\newcommand{\ba}{\begin{array}}
\newcommand{\ea}{\end{array}}
\newcommand{\beqa}{\begin{eqnarray}}
\newcommand{\eeqa}{\end{eqnarray}}
\newcommand{\beqas}{\begin{eqnarray*}}
\newcommand{\eeqas}{\end{eqnarray*}}
\newcommand{\bfg}{\begin{figure}}
\newcommand{\efg}{\end{figure}}
\newcommand{\bds}{\begin{displaymath}}
\newcommand{\eds}{\end{displaymath}}
\newcommand{\btb}{\begin{tabbing}}
\newcommand{\etb}{\end{tabbing}}
\bc {
\huge  Dirac Equation on the Torus and Rationally Extended Trigonometric Potentials within Supersymmetric QM }  \ec

\vs{1cm}

\bc
{\it  \"Ozlem Ye\c{s}ilta\c{s}$^{*}${\footnote {e-mail : yesiltas@gazi.edu.tr}   \\
$^{*}$Department of Physics, Faculty of Science, Gazi University,
06500 Ankara, Turkey\\
\vspace{.16cm}

}} \ec \vs{1cm}

\begin{abstract}
The exact solutions of the $(2+1)$ dimensional Dirac equation on the torus and the new extension and generalization of the trigonometric P\"{o}schl-Teller potential families in terms of the torus parameters are obtained. Supersymmetric quantum mechanical techniques are used to get the extended potentials when the inner and outer radii of the torus are both equal and inequal. In addition, using the aspects of the Lie algebraic approaches, the $iso(2,1)$ algebra is also applied to the system  where we have arrived at the spectrum solutions of the extended potentials using the Casimir operator that matches with the results of the exact solutions.
\end{abstract}
\noindent {\bf keyword:}   Dirac equation, curved space-time, torus \\


\section{Introduction}
There are several potentials which are  used in both theoretical and applied  physics  with exact solutions in non-relativistic quantum mechanics. These potential classes can also give physical results in the successful union of quantum mechanics and special relativity where the Dirac equation has a successfully explanation on the antimatter, spin, and the realistic behavior of atoms \cite{Dirac}. On the other hand, the gravitational field effects on  some quantum mechanical systems have been studied as an exciting research field \cite{par}, \cite{bez}, \cite{fins}. From the symmetries of the Dirac equation \cite{sym}, Hermiticity and uniqueness \cite{nez}, to the factorization and pseudo-supersymmetry \cite{ir}, \cite{yes}, covariant form of the Dirac equation and its aspects are studied. The gravitational background can bring some mathematical difficulties through some geometries. One of them is the solutions of the wave equations on the torus geometry \cite{ribbon}, \cite{car}, \cite{can}, \cite{aust}. At the same time, in physical applications, such as graphene related ones, it is stated that the curvature of the material can change the electron density of the states. In \cite{ribbon}, graphene nanoribbon along the surface of a torus is examined within the long wave approximation where the Dirac equation is solved approximately. Then, the recent study aims to bring a new viewpoint to the problem of the exact solutions for the Dirac equation on the torus which may lead to both applied and theoretical interest in the recent studies. When considering the works which are about the general theory of relativity and the quantum mechanics unification, there is a fact that the curvature of space-time at the position of the atom can affect the spectrum. Therefore, the problem of electron and its perturbed energy spectrum owing to the gravitational field can bring more problems in the solutions through the geometries of the space-time. To our knowledge, there are a few studies about the torus parametrization of the wave equations and exact solutions, then, we have devoted our motivation to find soluble potential models. In this paper, Section I involves the  relativistic quantum mechanical wave equation in the gravitational background for a massless fermion where the Fermi velocity is
taken as a position dependent function \cite{omar}. Supersymmetric partner potentials are obtained for the transformed Dirac Hamiltonian. The spectrum and spinor solutions are given when the  inner and outer radii of the torus are equal and inequal. In section III, we present the $iso(2,1)$ Lie algebraic computations for the
Dirac Hamiltonians.

\section{Dirac equation on the torus}
Formulating the Dirac equation in curved spacetime, it is known that all  metrics related by a general coordinate transformation are physically equivalent
and  physical observables in gravitation field should be invariant under general coordinate transformations. This  is known as general covariance principle which necessities transforming a tensor in the flat spacetime as a tensor under general transformations in the curved  manifold. The covariant generalization
 of the Dirac equation to curved space was independently introduced  by Weyl  and by Fock  \cite{weyl}, \cite{fock}. Then, the Dirac equation
can be written in terms of vierbein fields and gravitational spin connection as \cite{weyl}
\begin{equation}\label{1}
  [i\gamma^{\mu}(\partial_{\mu}-\Gamma_{\mu})]\Psi=0
\end{equation}
where $\Gamma_{\mu}$ is the spin connection.  And  $\Psi$ is the spinor  which includes electron's wave-functions $\Psi=\left(
                                                                                                                 \begin{array}{c}
                                                                                                                   \psi_1 \\
                                                                                                                   \psi_2 \\
                                                                                                                 \end{array}
                                                                                                               \right)$
near the Dirac point. The Dirac matrices $\gamma^{\mu}$ in curved spacetime satisfy,
\begin{equation}\label{2}
  \{\gamma^{\mu}, \gamma^{\nu}\}=2g^{\mu \nu}.
\end{equation}
Here $g^{\mu\nu}$ is the metric tensor and the tetrad(vierbein) frames field is defined as
\begin{equation}\label{3}
    g_{\mu \nu}=e^{a}_{\mu}e^{b}_{\nu}\eta_{ab}
\end{equation}
where $\eta_{ab}=diag[1 ~~ -1 ~~-1]$. The metric  for the torus surface is given by
\begin{equation}\label{4}
    ds^{2}=dt^{2}-a^{2}dv^{2}-(c+a \cos v)^{2}du^{2}.
\end{equation}
In the metric given above,  the inner radius of the torus is $c$, the outer radius is shown by $a$ and $u, v \in [0, 2\pi )$. We can use the spin connection
formula which is
\begin{equation}\label{5}
    \Gamma_{\mu}=\frac{1}{4}\gamma_{a}\gamma_{b}e_{\lambda}^{a}g^{\lambda\sigma}(\partial_{\mu}e^{b}_{\sigma}-\Gamma^{\lambda}_{\mu\sigma}e_{\lambda}^{b})
\end{equation}
where $\Gamma^{\lambda}_{\mu\sigma}$ is the Christoffel symbols. In \cite{ribbon}, the symbols $\Gamma^{\lambda}_{\mu\sigma}$ were given in terms of the variable
$R(x_{1})$ which corresponds to $R(v)$ in our work. So one can obtain them  as,
\begin{equation}\label{6}
    \Gamma^{2}_{12}=-\frac{a\sin v}{c+a\cos v},~~ \Gamma^{1}_{22}=\frac{1}{a}(c+a\cos v)\sin v
\end{equation}
and accordingly,
\begin{equation}\label{7}
    \Gamma_0=0,~~~\Gamma_1=0,~~~\Gamma_2=-\gamma_{1}\gamma_2\frac{a \sin v}{2(c+a\cos v)}.
\end{equation}
For the Dirac matrices we will use $\sigma_{j}$ Pauli matrices as $\gamma_0=\sigma_3$, $\gamma_1=-i\sigma_2$, $\gamma_2=-i\sigma_1$.
We use (\ref{6}) and (\ref{7}) in (\ref{1}) and we get \cite{ribbon}
\begin{eqnarray}
 ( V^{-1}_{F}\partial_{t}) \psi_1 &=& (-\frac{1}{a^{2}}\partial_{v}-\frac{i}{R^{2}}\partial_{u}+\frac{R^{'}}{2R^{2}a})\psi_2 \\
 ( V^{-1}_{F}\partial_{t})  \psi_2 &=& (-\frac{1}{a^{2}}\partial_{v}+\frac{i}{R^{2}}\partial_{u}+\frac{R^{'}}{2R^{2}a})\psi_1,
\end{eqnarray}
where $R=c+a\cos v$. We note that $\partial_{0}=V^{-1}_{F}\partial_{t}$ is taken and the Fermi velocity $V^{-1}_{F}$ may not be a constant \cite{omar}. Using $\left(    \begin{array}{c}
                                                                                                                   \psi_1(t,v,u) \\
                                                                                                                   \psi_2(t, v,u) \\
                                                                                                                 \end{array}
                                                                                                               \right)=e^{iEt-iku}\left(
                                                                                                                 \begin{array}{c}
                                                                                                                   \psi_1(v) \\
                                                                                                                   \psi_2(v) \\
                                                                                                                 \end{array}
                                                                                                               \right)$, we obtain
\begin{equation}\label{8}
\begin{split}
  -\frac{d^{2}\psi_1}{dx^{2}}+\frac{aV_{F}(x)R'(x)-R(x)^{2}V'_{F}(x)}{V_{F}(x)R(x)^{2}}\frac{d\psi_1}{dx}+
  (\frac{k^{2}a^{4}}{R^{4}(x)}-\frac{2kR'(x)a^{2}}{R(x)^{3}}-\frac{R'(x)^{2}a^{2}}{4R(x)^{4}}-\frac{R'(x)^{2}a}{R(x)^{3}}+\frac{kV'_{F}(x) a^{2}}{R(x)^{2}V_{F}(x)} & +\\ \frac{R'(x) V'_{F}(x)a}{2R(x)^{2}V_{F}(x)}+\frac{R''(x)a}{2R(x)^{2}}
  )\psi_1(x)=\frac{E^{2}a^{4}}{V_{F}(x)^{2}}\psi_1(x)
  \end{split}
\end{equation}

\begin{equation}\label{9}
\begin{split}
  -\frac{d^{2}\psi_2}{dx^{2}}+\frac{aV_{F}(x)R'(x)-R(x)^{2}V'_{F}(x)}{V_{F}(x)R(x)^{2}}\frac{d\psi_2}{dx}+
  (\frac{k^{2}a^{4}}{R^{4}(x)}+\frac{2kR'(x)a^{2}}{R(x)^{3}}-\frac{R'(x)^{2}a^{2}}{4R(x)^{4}}-\frac{R'(x)^{2}a}{R(x)^{3}}-\frac{kV'_{F}(x) a^{2}}{R(x)^{2}V_{F}(x)} & +\\ \frac{R'(x) V'_{F}(x)a}{2R(x)^{2}V_{F}(x)}+\frac{R''(x)a}{2R(x)^{2}}
  )\psi_2(x)=\frac{E^{2}a^{4}}{V_{F}(x)^{2}} \psi_2(x)
  \end{split}
\end{equation}
where we use $v \rightarrow x $ for the sake of simplicity. Considering (\ref{8}), the solutions may take the form
\begin{equation}\label{10}
  \psi_1(x)=f(x)F_{1}(g(x)).
\end{equation}
Then we have
\begin{equation}\label{11}
  F_{1}''(x)+\left(\frac{2f'}{fg'}+\frac{g''}{g'^{2}}+\frac{V'_F}{V_F g'}-a\frac{R'}{R^{2}g'}\right)F_{1}'(x)+\left(\frac{E^{2}a^{4}/V^{2}_F-U_{1}(x)}{g'^{2}}+
  \frac{f'}{fg'^{2}}(\frac{V'_F}{V_F}-\frac{aR'}{R^{2}})+\frac{f''}{fg'^{2}}\right)F_{1}(x)=0.
\end{equation}
(\ref{11}) can also be expressed as
\begin{equation}\label{12}
  F_{1}''(g)+\textbf{Q(g)}F_{1}'(g)+\textbf{R(g)}F_{1}(g)=0,
\end{equation}
where
\begin{eqnarray}
  \textbf{Q(g(x))} &=& \frac{2f'}{fg'}+\frac{g''}{g'^{2}}+\frac{V'_F}{V_F g'}-a\frac{R'}{R^{2}g'} \\
  \textbf{R(g(x))} &=&  \frac{E^{2}a^{4}/V^{2}_F-U_{1}(x)}{g'^{2}}+
  \frac{f'}{fg'^{2}}(\frac{V'_F}{V_F}-\frac{aR'}{R^{2}})+\frac{f''}{fg'^{2}}.
\end{eqnarray}
We may give $U(x)$ as below
\begin{equation}\label{13}
  U_{1}(x)=\frac{k^{2}a^{4}}{R^{4}}-\frac{2kR'(x)a^{2}}{R(x)^{3}}-\frac{R'(x)^{2}a^{2}}{4R(x)^{4}}-\frac{R'(x)^{2}a}{R(x)^{3}}+\frac{kV'_{F}(x) a^{2}}{R(x)^{2}V_{F}(x)}  + \frac{R'(x) V'_{F}(x)a}{2R(x)^{2}V_{F}(x)}+\frac{R''(x)a}{2R(x)^{2}}.
\end{equation}
One can make the coefficient of $F_{1}'(g)$ as zero and $f(x)$ can be found as
\begin{equation}\label{14}
  f(x)=\frac{C_1 e^{-\frac{a}{2R(x)}}}{\sqrt{g'(x)V_F(x)}}.
\end{equation}
Then, using $V_F(x)=\frac{1}{g'(x)}$ in (\ref{14}), (\ref{12}) becomes
\begin{equation}\label{15}
  F_{1}''(x)+(a^{4}E^{2}-V_{1}(x))F_{1}(x)=0,
\end{equation}
where
 \begin{equation}\label{16}
 V_{1}(x)=\frac{a^{4}k^{2}}{R^{4}(x)g'^{2}}-\frac{2a^{2}kR'(x)}{R^{3}(x)g'^{2}}-\frac{a^{2}kg''}{R^{2}(x)g'^{3}}.
 \end{equation}
In this work  \cite{bq}, it is shown that if a superpotential $W(x)$ which is a hyperbolic function
\begin{equation}\label{17}
  W(x)= A \coth x+ B \csc h x-\frac{\sinh x}{\cosh x+C},
\end{equation}
then, the pair of potentials are given as
\begin{eqnarray}\label{170}
  V_{1}^{(+)}(x) &=& (A(A+1)+B^{2})\csc h^{2}x+(2A+1)B \csc h x \coth x+\frac{(2A-1)C-2B}{C+\cosh x}+(A-1)^{2}+E \\
  V_{1}^{(-)}(x) &=&  (A(A-1)+B^{2})\csc h^{2}x+(2A-1)B \csc h x \coth x+\frac{(2A-3)C-2B}{C+\cosh x}+\frac{2(C^{2}-1)}{(C+\cosh x)^{2}}+(A-1)^{2}+E \nonumber
\end{eqnarray}
The authors removed the rational term in $V^{+}(x)$ using a parameter condition $C=\frac{2B}{2A-1}$. In this case, the potential $V^{+}(x)$ is known as generalized P\"{o}schl-Teller potential in the literature with the exact solutions. Now we will discuss the cases depending on the different choices of rational term in $W(x)$ superpotential. Now we shall look at (\ref{16}) to arrive at a solvable model given below. If we take
\begin{equation}\label{171}
  V_{1}(x)=\frac{a^{4}k^{2}}{R^{4}(x)g'^{2}}-\frac{2a^{2}kR'(x)}{R^{3}(x)g'^{2}}-\frac{a^{2}kg''}{R^{2}(x)g'^{3}}=V_1 \csc^{2} x +V_2\cot x \csc x+\varepsilon,
\end{equation}
where $V_1, V_2, \varepsilon$ are real constants. Hence, our function $g(x)$ satisfying (\ref{171}) can be found as
\begin{equation}\label{172}
  g(x)= C_1 \pm \int^{x}\frac{i \exp(\int^{y} \frac{a^{2}k-2R(z)R'(z)}{R^{2}(z)}dz)}{\sqrt{\mu(y)-C_2}}dy,
\end{equation}
where $C_1, C_2$ are constants and
\begin{equation}\label{173}
  \mu(y)=2\int^{y}-\frac{\exp(2\int^{t} \frac{a^{2}k-2R(z)R'(z)}{R^{2}(z)}dz)(2V_1+\varepsilon+V_2 \cos t+\varepsilon \cos 2t \csc^{2}t R^{2}(t))}{2a^{2}k}dt.
\end{equation}
Now we will search for the solutions for our systems in case of the equal and unequal torus radii.
%%%%%%%%%%%%%%%%%%%%%%%%%%%%%%%%%%%%%%%%%%%%%%%%%%%%%%%%%%%%%%%%%%%%%%%%%%%%%%%%%%%%%%%%%%%%%%%%%%%%%%%%%%%%%%%%%%%%%%%%%%%%%%%%%%%%%%%%%%%%%%%%%%%%%%%%%%%%%%%%%%%%%%%%%%%%%%%%%%%%%%%%%%%%%%%%%%%%%%%%%%%%%%%%%%%%%%%%%%%%%%%%%%
%%%%%%%%%%%%%%%%%%%%%%%%%%%%%%%%%%%%%%%%%%%%%%%%%%%%%%%%%%%%%%%%%%%%%%%%%%%%%%%%%%%%%%%%%%%%%%%%%%%%%%%%%%%%%%%%%%%%%%%%%%%%%%%%%%%%%%%%%%%%%%%%%%%%%%%%%%%%%%%%%%%%%%%%%%%%%%%%%%%%%%%%%%%%%%%%%%%%%%%%%%%%%%%%%%%%%%%%%%%%%%%%%%%
\subsection{ Equal inner an outer radii; $a=c$}
First we assume that the superpotential $W_1(x)$ is a trigonometric function which is
\begin{equation}\label{19}
  W_1(x)=A\cot x+B\csc x+\frac{\lambda \sin x}{c+a\cos x}
\end{equation}
and we get the partner potentials
\begin{equation}\label{20}
  V^{-}_{1}(x)=W^{2}_{1}(x)-W'_{1}(x)\\
  =(1+2A)B\cot x \csc x+(A+B^{2}+A^{2})\csc^{2}x-A^{2}+\frac{\lambda(-a+\lambda)\sin^{2}x}{(c+a\cos x)^{2}}+\frac{\lambda(2B+(2A-1)\cos x)}{c+a\cos x}
\end{equation}
\begin{equation}\label{21}
  V^{+}_{1}(x)=W^{2}_{2}(x)+W'_{2}(x)\\
  =(2A-1)B\cot x \csc x+(B^{2}-A+A^{2})\csc^{2}x-A^{2}+\frac{\lambda(a+\lambda)\sin^{2}x}{(c+a\cos x)^{2}}+\frac{\lambda(2B+(2A+1)\cos x)}{c+a\cos x}.
\end{equation}
It can be seen from (\ref{20}) that in case of an applied special condition $a=1, \lambda=1$ and $x\rightarrow i x$, one can obtain the system (\ref{170}).
In order to find a parameter condition on the rational term, we can simplify (\ref{20}) as
\begin{equation}\label{22}
\begin{split}
    V^{-}_{1}(x)=(1+2A)B\cot x \csc x+(A+B^{2}+A^{2})\csc^{2}x-A^{2}\\+\lambda\frac{2a(A-1)+4Bc+\lambda+(4aB-2c+4Ac)\cos x+(2aA-\lambda)\cos 2x}{2(c+a\cos x)^{2}}.
\end{split}
\end{equation}
Then, we can find this conditions as
\begin{eqnarray}\label{23}
  \lambda &=& 2aA \\
  c &=& \frac{2aB}{1-2A} \\
  A &=& \frac{1}{2}(1 \pm 2B)\\
  a &=& |c|.
\end{eqnarray}
 Thus we have,
\begin{equation}\label{24}
  V^{-}_{1}(x)=(1+2A)B\cot x \csc x+(A+B^{2}+A^{2})\csc^{2}x-A^{2},
\end{equation}
which is known as trigonometric P\"{o}schl-Teller potential. On the other hand, we can give the partner potential $V_{+}(x)$ as
\begin{equation}\label{25}
V^{+}_{1}(x)= 2B^{2}\cot x \csc x+(B^{2}+A(A-1))\csc^{2} x+\frac{\lambda(a+\lambda)\sin^{2}x}{(a+a\cos x)^{2}}+\frac{\lambda(1-2A+(2A+1)\cos x)}{a+a\cos x}-A^{2}
\end{equation}
Here, it is  well known that  (\ref{24}) and (\ref{25}) are isospectral partner potentials except the ground-state. Comparing (\ref{24}) and $V_{1}(x)$ in (\ref{171}), we obtain
\begin{equation}\label{26}
  V_{1}(x)=V^{-}_{1}(x),~~~~V_1=(1+2A)B,~~~~V_2=A(A+1)+B^{2}, ~~~~\varepsilon=-A^{2}.
\end{equation}
The solutions of $V_{-}(x)$ are already known as \cite{levai}
\begin{equation}\label{27}
  E^{-}_{1,n}=\pm \frac{\sqrt{(n-\frac{\lambda}{2a})^{2}-\frac{\lambda^{2}}{4a^{2}}}}{a^{2}}.
\end{equation}
This result shows that when $a\rightarrow \infty$, $E\rightarrow 0$. Reminding the eigenvalue equation (\ref{15}), we can give
\begin{equation}\label{28}
  \hat{H}_{-}F^{(-)}_{1,n}(x)=E^{2}a^{4}F^{(-)}_{1,n}(x), ~~\hat{H}_{-}=\hat{A}^{\dag}\hat{A},~~~~\hat{A}=\frac{d}{dx}+W_{1}(x). \end{equation}
Then, the solutions are written as \cite{levai}
\begin{equation}\label{29}
  F^{(-)}_{1,n}(x) \sim (1-\cos x)^{\frac{-A-B}{2}}(1+\cos x)^{\frac{-A+B}{2}}P^{(-A-B-1/2,-A+B-1/2)}_{n}.
\end{equation}
In fact, the complete solutions are not given yet. The spinor solutions can be given as
\begin{equation}\label{30}
  \psi_1,n(x)= N_{1} \exp(-\frac{a}{2(a+a\cos x)})(1-\cos x)^{\frac{-A-B}{2}}(1+\cos x)^{\frac{-A+B}{2}}P^{(-A-B-1/2,-A+B-1/2)}_{n}(\cos x),
\end{equation}
where $N_1$ is the normalization constant. The solutions $F_{+}(x)$ corresponding the ones for the partner potential $V_{1,+}(x)$ can be found. The system can be summarized as follows
\begin{equation}\label{31}
   \hat{H}_{+}F^{+}_{1}(x)=E^{2}a^{4}F^{+}_{1}(x), ~~H_{+}=\hat{A}\hat{A}^{\dag}.
\end{equation}
For (\ref{25}),
\begin{equation}\label{32}
 F^{(+)}_{1,n}=\frac{1}{\sqrt{E_{n,-}}}\hat{A}F^{(-)}_{1,n}
\end{equation}
and it is reminded that the energy states of these Hamiltonians are given by
\begin{equation}\label{33}
  E^{-}_{1,n+1}=E^{+}_{2,n}.
\end{equation}
At the same time, we can find $F^{(+)}_{1,n}(x)$ solutions using (\ref{32}),
\begin{equation}\label{farti}
\begin{split}
  F^{(+)}_{1,n}(x)\sim \frac{(1-\cos x)^{\frac{-A-B}{2}}(1+\cos x)^{\frac{1}{2}(B-A)}}{a}(\frac{a(2A-n)}{2}P^{(1/2-A-B, 1/2-A+B)}_{n-1}(\cos x)\sin x+\\
  \lambda P^{(-1/2-A-B, -1/2-A+B)}_{n}(\cos x)\tan \frac{x}{2}).
  \end{split}
\end{equation}
When it comes to the solutions $\psi_2(x)$ of (\ref{9}) which shares the same energy with (\ref{8}), we give
\begin{equation}\label{34}
    U_{2}(x)=\frac{k^{2}a^{4}}{R^{4}}\textcolor[rgb]{1.00,0.00,0.00}{+}\frac{2kR'(x)a^{2}}{R(x)^{3}}-\frac{R'(x)^{2}a^{2}}{4R(x)^{4}}-
    \frac{R'(x)^{2}a}{R(x)^{3}}\textcolor[rgb]{1.00,0.00,0.00}{-}\frac{kV'_{F}(x) a^{2}}{R(x)^{2}V_{F}(x)}  + \frac{R'(x) V'_{F}(x)a}{2R(x)^{2}V_{F}(x)}+\frac{R''(x)a}{2R(x)^{2}}.
\end{equation}
The wavefunction mapping can be given as
\begin{equation}\label{35}
  \psi_2(x)=f(x)F_{2}(g_{2}(x))
\end{equation}
and we obtain,
\begin{equation}\label{36}
  F_{2}''(x)+(a^{4}E^{2}-V_{2}(x))F_{2}(x)=0,~~~~V_F(x)=\frac{1}{g_{2}'(x)}
\end{equation}
where
 \begin{equation}\label{37}
 V_{2}(x)=\frac{a^{4}k^{2}}{R^{4}(x)g_{2}'^{2}}+\frac{2a^{2}kR'(x)}{R^{3}(x)g_{2}'^{2}}+\frac{a^{2}kg_{2}''}{R^{2}(x)g_{2}'^{3}}.
\end{equation}
 Then, we can give $V_2(x)$ as
\begin{equation}\label{38}
  V_2(x)  = \varepsilon+P_2  \cot x \csc x+P_1 \csc^{2} x,
\end{equation}
$P_1$ and $P_2$ are real constants and equating (\ref{37}) and (\ref{38}), $g_{2}(x)$ is found to be
\begin{equation}\label{39}
  g_2(x)= C_2\pm \int^{x} \frac{i\exp(-\int^{u}\frac{a^{2}k+2R(y)R'(y)}{R^{2}(y)}dy)}{\sqrt{-C_1+2\int^{u}\frac{\exp(-2\int^{y}\frac{a^{2}k+2R(y)R'(y)}{R^{2}(y)}dy)(2P_1+\varepsilon+2P_2\cos z-\varepsilon \cos 2z)R^{2}(z)\csc^{2}z}{2a^{2}k} dz}}du.
\end{equation}
If the parameters are chosen as $P_1=V_2$ and $P_2=-V_1$, then we get
\begin{equation}\label{40}
  V_2(x)  = -A^{2}-B(1+2A)  \cot x \csc x+(A(A+1)+B^{2}) \csc^{2} x,
\end{equation}
and using (\ref{23}), $P_1$ and $P_2$ can be expressed in terms of the torus parameter and hence, one can give the supersymmetric partner potentials for the $V_{2}(x)$ which are $V^{\pm}_{2}$ are
\begin{equation}\label{41}
  V^{-}_{2}(x) = V_2(x)=-\frac{\lambda^{2}}{4a^{2}}-(\frac{1}{2}-\frac{\lambda^{2}}{2a^{2}})\cot x \csc x+\frac{1}{2}(\frac{1}{2}+\frac{\lambda^{2}}{a^{2}})\csc^{2}x
\end{equation}
and
\begin{equation}\label{41}
\begin{split}
  V^{+}_{2}(x) =  2B^{2}\cot x \csc x+(B^{2}+A(A-1))\csc^{2} x+(1-2B)\csc^{2}\frac{x}{2}-A^{2}.
\end{split}
\end{equation}
\begin{equation}\nonumber
~~~~~~~~~~=-\frac{\lambda^{2}}{4a^{2}}+\frac{1}{2}(1-\frac{\lambda}{a})^{2}\cot x\csc x+(\frac{1}{4}+ \frac{\lambda(\lambda-2)}{2a^{2}})\csc^{2}x+\frac{\lambda}{a}\csc^{2}\frac{x}{2}
\end{equation}

Now, for the system (\ref{9}) we can give the solutions given below
\begin{equation}\label{42}
\psi_2,n(x)= N_{2} \exp(-\frac{a}{2(a+a\cos x)})(1-\cos x)^{(a-2\lambda)/4a}(1+\cos x)^{-\frac{1}{4}}P^{(-1,-\lambda/a)}_{n}(\cos x).
\end{equation}
%%%%%%%%%%%%%%%%%%%%%%%%%%%%%%%%%%%%%%%%%%%%%%%%%%%%%%%%%%%%%%%%%%%%%%%%%%%%%%%%%%%%%%%%%%%%%%%%%%%%%%%%%%%%%%%%%%%%%%%%%%%%%%%%%%%%%%%%%%%%%%%%%%%%%%%%%%%%%%%%%%%%%%%%%%%%%%%%%%%%%%%%%%%%%%%%%%%%%%%%%%%%%%%%%%%%%%%%%%%%%%%
%%%%%%%%%%%%%%%%%%%%%%%%%%%%%%%%%%%%%%%%%%%%%%%%%%%%%%%%%%%%%%%%%%%%%%%%%%%%%%%%%%%%%%%%%%%%%%%%%%%%%%%%%%%%%%%%%%%%%%%%%%%%%%%%%%%%%%%%%%%%%%%%%%%%%%%%%%%%%%%%%%%%%%%%%%%%%%%%%%%%%%%%%%%%%%%%%%%%%%%%%%%%%%%%%%%%%%%%%%%%%%%%
\subsection{different inner and outer radii;  $a\neq c$}
\begin{equation}\label{43}
  W_{2}(x)=A\cot x+B\csc x+\frac{G(x)}{c+a\cos x}
\end{equation}
where $G(x)$ is the unknown function and we get
\begin{equation}\label{44}
\begin{split}
  V^{-}_{1}(x)=B(1+2A)\cot x \csc x+(A^{2}+A+B^{2})\csc^{2}x-A^{2}+\\
  \frac{G(x)^{2}+(2(c+a\cos x)(B+A\cos x)\csc x-a\sin x)G(x)+(c+a\cos x)G'(x)}{(c+a\cos x)^{2}}.
\end{split}
\end{equation}
 In order to make the rational term zero, $G(x)$ can be found as   below
\begin{equation}\label{45}
  G(x)=\frac{c+a\cos x}{(\sin x)^{-2A+2B}(1-\cos x)^{-2B}(4^{-A}+4^{B}\mathcal{B}(\cos^{2}\frac{x}{2});\frac{1}{2}+A-B,\frac{1}{2}+A+B)},
\end{equation}
where $\mathcal{B}(z;s,w)$ is the incomplete beta function which is equal to
\begin{equation}\label{46}
  \mathcal{B}(z;s,w)=\int^{z} u^{s-1}(1-u)^{w-1} du,~~~~~~z\in (0,1)
\end{equation}
and the ordinary(complete) beta function is $\mathcal{B}(s,w)=\mathcal{B}(1;s,w)$. Then, the partner potentials are obtained as
\begin{equation}\label{47}
  V^{-}_{1}(x)=B(1+2A)\cot x \csc x+(A^{2}+A+B^{2})\csc^{2}x-A^{2}
\end{equation}
and
\begin{equation}\label{48}
\begin{split}
  V^{+}_{1}(x)=(2A-1)B\cot x \csc x+(A^{2}-A+B^{2})\csc^{2}x-A^{2}+\\
  \frac{(\mathcal{B}(\cos^{2}x/2,1/2+A-B,1/2+A+B)\xi(x)+\chi(x))(\sin\frac{x}{2})^{-1+2A+2B}}{2^{2A+2B}\mathcal{B}(\cos^{2}x/2,1/2+A-B,1/2+A+B)(\sin x)^{2A+2B}+2^{-2A-2B}C_1( \sin\frac{x}{2})^{2A+2B}(\sin\frac{x}{2})^{2A+2B}}
  \end{split}
\end{equation}
where
\begin{eqnarray}
  \xi(x) &=& 2^{3/2+8A}(\cos\frac{x}{2})^{6A+2B}\sqrt{1-\cos x}(B+A\cos x)\csc x (\sin\frac{x}{2})^{4A+4B} \\
  \chi(x) &=& 2^{-1/2+4A}(\cos\frac{x}{2})^{6A}(\sin\frac{x}{2})^{4A+4B}(4C_1 (\cos\frac{x}{2})^{2B}\sqrt{1-\cos x}(B+A\cos x)\csc x +\\ & &16^{A}(\cos\frac{x}{2})^{2A}(\sin\frac{x}{2})^{2A+2B}(\sqrt{1-\cos x}+\sqrt{1+\cos x}\tan\frac{x}{2})). \nonumber
\end{eqnarray}
For the unknown $G(x)$ case, there are no restrictions for the parameters. (\ref{47}) and (\ref{48}) share the same spectrum except the ground state which is
\begin{equation}\label{4790}
    E^{-}_{1,n}=\pm \frac{\sqrt{(n-A)^{2}-A^{2}}}{a^{2}}.
\end{equation}
On the other hand, we can give an ansatz for the $W(x)$,
\begin{equation}\label{480}
  W(x)=A\cot x+B\csc x+\frac{G(x)+\lambda \sin x}{c+a\cos x}, \end{equation}
\begin{eqnarray}\label{49}
  V^{-}_{1}(x)&=& B(2A+1)\cot x \csc x+(A^{2}+A+B^{2})\csc^{2}x-A^{2}+ \\
  & &   \frac{\lambda(2a(A-1)+4Bc+\lambda)+2\lambda\cos x(2aB+c(2A-1))+(2aA-\lambda)\cos 2x+\mathcal{G}(x)}{2(c+a\cos x)^{2}}, \nonumber
  \end{eqnarray}
where
\begin{equation}\label{50}
  \mathcal{G}(x)=2G(x)^{2}+G(x)(4aB+4Ac)\cot x+4aA\cos x\cot x+4Bc\csc x+(4\lambda-2a)\sin x-2(c+a\cos x)G'(x).
\end{equation}
If we terminate the rational term in (\ref{49}), we find
\begin{equation}\label{51}
  1+2a(A-1)+4Bc+\lambda = 0,~~~~ 4aB-2c+4Ac = 0,~~~~ 2aA-\lambda = 0,~~~~ \mathcal{G}(x) = 0, \end{equation}
and we can also obtain
\begin{equation}\label{52}
  A = \frac{\lambda}{2A},~~~~ B =\frac{1}{2}( \pm \frac{\sqrt{1-2a+2\lambda}}{\sqrt{-2a+2\lambda}}-\frac{\lambda\sqrt{1-2a+2\lambda}}{a\sqrt{-2a+2\lambda}}),~~~~ c= \pm \frac{a\sqrt{1-2a+2\lambda}}{\sqrt{-2a+2\lambda}}.
\end{equation}
To terminate the remaining function on the nominator of (\ref{49}), one can take $\mathcal{G}(x)=0$, hence  $G(x)$ is obtained  as
\begin{equation}\label{53}
G(x)= \frac{ 8^{A}(1+\cos x)^{A}(c+a\cos x)}{C_1 (\cot\frac{x}{2})^{2B}(c+a\cos x)^{2\lambda/a}(\sin\frac{x}{2})^{-2A}\varrho(x)}
\end{equation}
where
\begin{equation}\label{54}
\begin{split}
 \varrho(x)= -\frac{1}{1+2A+2B}2^{4A+1/2} Appel F_{1}(1/2+A+B,1/2-A+B,\frac{2\lambda}{a},3/2+A+B,\sin^{2}x/2,\frac{a-a\cos x}{a+c})\\ (\cos\frac{x}{2})^{2B}\sqrt{1+\cos x}
 (\frac{c+a\cos x}{a+c})^{\frac{2\lambda}{a}}\tan\frac{x}{2}.
 \end{split}
\end{equation}
Here, $F_{1}(\alpha,\beta,\beta',\gamma;x;y)$ is the Appel hypergeometric function with two variables.  Thus, $V_{1,-}(x)$ can be given by (\ref{47}) and $V_{+}(x)$ becomes
\begin{equation}\label{55}
\begin{split}
  V^{+}_{1}(x)=(2A-1)B\cot x\csc x+(B^{2}+A^{2}-A)\csc^{2}x-A^{2}+\\
  \lambda\frac{2a(A+1)+4Bc+\lambda+2(2aB+c+2Ac)\cos x+(2aA-\lambda)\cos 2x}{2(c+a\cos x)^{2}}+\\ \frac{\chi(x)}{(c+a\cos x)(C_1(\cot\frac{x}{2})^{2B}(c+a\cos x)^{\frac{2\lambda}{a}})-\varrho(x)}, \end{split}
\end{equation}
where
\begin{equation}\label{56}
\begin{split}
  \chi(x)=(8^{\lambda}((1+\cos x)^{A} \frac{2\varrho(x)(aA+2Bc+\lambda+2(aB+Ac)\cos x+(aA-\lambda)\cos 2x)}{\sqrt{1+\cos x}}+\\ 2(8^{A}(1+\cos x)^{A})(c+a\cos x)+\\ C_1 (\cos\frac{x}{2})^{2B}(c+a\cos x)^{\frac{2\lambda}{a}}(aA+2Bc+\lambda+2(aB+Ac)\cos x+(aA-\lambda)\cos 2x)\csc x (\sin\frac{x}{2})^{-2A-2B})).
\end{split}
\end{equation}
Similarly, one can obtain the partner potentials for the system given in (\ref{9}).
%%%%%%%%%%%%%%%%%%%%%%%%%%%%%%%%%%%%%%%%%%%%%%%%%%%%%%%%%%%%%%%%%%%%%%%%%%%%%%%%%%%%%%%%%%%%%%%%%%%%%%%%%%%%%%%%%%%%%%%%%%%%%%%%%%%%%%%%%%%%%%%%%%%%%%%%
%%%%%%%%%%%%%%%%%%%%%%%%%%%%%%%%%%%%%%%%%%%%%%%%%%%%%%%%%%%%%%%%%%%%%%%%%%%%%%%%%%%%%%%%%%%%%%%%%%%%%%%%%%%%%%%%%%%%%%%%%%%%%%%%%%%%%%%%%%%%%%%%%%%%%%%%%%%
\section{The $iso(2,1)$ algebraic approach}
Let us look at the  operators given below
\begin{equation}\label{57}
  J_{\pm}=i e^{\pm i\phi}(\pm \frac{\partial}{\partial x}-((-i\frac{\partial}{\partial x}\pm \frac{1}{2})S(x)-T(x))+U(x,\mu \pm 1/2))
\end{equation}
and
\begin{equation}\label{58}
  J_3=-i\frac{\partial}{\partial \phi}.
\end{equation}
Here, the term $U(x,\mu \pm 1/2)$ is the modification operator which is used in \cite{yadav}, $\mu$ is a constant. Without $U(x,\mu \pm 1/2)$, these operators are known as  those given in the $iso(2,1)$ algebra but this functional operator, $U(x,\mu \pm 1/2)$, helps to construct the algebra for the rationally extended potentials. Here, we discuss the extended trigonometric P\"{o}schl-Teller potential within the $iso(2,1)$ algebra. These operators provide the
commutation relations which are given as
\begin{equation}\label{59}
  [J_{+}, J_{-}]=-2J_{3},~~~~[J_{3}, J_{\pm}]=\pm J_{\pm}.
\end{equation}
If $J_{\pm}$ and $J_{3}$ are used in (\ref{59}), then, one can get
\begin{equation}\label{60}
  S'(x)-S(x)^{2}=1,~~~~T'(x)-S(x)T(x)=0.
\end{equation}
The constraints in (\ref{59}) also lead to \cite{yadav}
\begin{equation}\label{61}
\begin{split}
  U_{1}(x)^{2}-\frac{d}{dx}U_{1}(x)+2U_{1}(x)(F(x)(\mu+\frac{1}{2})-G(x))-\\
  (U_{2}(x)^{2}-\frac{d}{dx}U_{2}(x)+2U_{2}(x)(F(x)\mu_1-G(x)))=0.
\end{split}
\end{equation}
In this study we will use $U(x,\mu+1/2)=U_{1}(x)$ and $U(x,\mu-1/2)=U_{2}(x)$.  And the Hamiltonian in (\ref{28}) can be given in terms of the Casimir operator $J^{2}=J^{2}_{3}-\frac{1}{2}(J_{+}J_{-}+J_{-}J_{+})$,  and we may denote the Hamiltonian using $J^{2}$ as
\begin{equation}\label{62}
 H=J^{2}+\frac{1}{4}.
\end{equation}
Here, the operators act on the physical states which are given by
\begin{equation}\label{63}
  J^{2}|j, \mu \rangle=j(j+1)|j, \mu \rangle
\end{equation}
\begin{equation}\label{64}
  J_{3}|j, \mu \rangle=\mu|j, \mu \rangle
\end{equation}
\begin{equation}\label{65}
  J_{\pm} |j, \mu \rangle=\sqrt{-(j \mp \mu)(j \pm \mu +1)}|j, \mu\pm 1 \rangle.
\end{equation}
And, $|j, \mu \rangle$ can be written in the function space as
\begin{equation}\label{66}
  |j, \mu \rangle=\psi_{j\mu}(x)e^{i\mu \phi}.
  \end{equation}
If we express $J^{2}$ using the operators as \cite{yadav}
\begin{equation}\label{jkare}
\begin{split}
  J^{2}=-\frac{d^{2}}{dx^{2}}+(1+F(x)^{2})(J^{2}_{3}-1/4)-2G'(x)(J_3)+G(x)^{2}-\frac{1}{4}+U_{1}^{2}(x)+((J_3+1/2)F(x)-\\ G(x))U_{1}(x)+
    U_{1}(x)((J_3+1/2)F(x)-G(x))-\frac{d}{dx}U_{1}(x)\\
    =-\frac{d^{2}}{dx^{2}}+V(x).
  \end{split}
\end{equation}
Choosing our functions $S(x)$ and $T(x)$ as
\begin{equation}\label{67}
  S(x)=-\cot x,~~~~T(x)=B_{1}\csc x,
\end{equation}
and using a suggestion for each   $U_{1}(x), U_2(x)$ which are given by
\begin{equation}\label{68}
  U_{1}(x)=-\frac{K_1 \sin x}{c+a\cos x},~~~~U_{~2}(x)=\frac{K_2 \sin x}{c+a\cos x}, \end{equation}
and using $U_1(x)$ in (\ref{jkare}), we can find the potential which is an element of $J^{2}$,
\begin{equation}\label{69}
  V(x)= -\frac{1}{4}+2B_1 \mu \cot x \csc x+(\mu^{2}+B^{2}_{1}-1/4)\csc^{2} x+\frac{2K_1 (B_1+(\mu+1)\cos x)}{c+a\cos x}+\frac{K_1(a+K_1)\sin^{2} x}{(c+a\cos x)^{2}}.
\end{equation}
For the $a=c$ case, we can compare (\ref{69}) and (\ref{20}), then we get,
\begin{equation}\label{70}
  \lambda=-K_1,~~~~B=-B_1,~~~~A=-\mu-1/2.
\end{equation}
Moreover, (\ref{61}) can be satisfied in case of chosen $U_{1}(x)$ and and $U_{2}(x)$ in  (\ref{68})  and the  parameter conditions as given below
\begin{equation}\label{71}
  B_1=-\frac{c+K_1}{2c},~~~~\mu=\frac{K_1}{2c}-\frac{1}{2},~~~~a=c,~~~~K_2=-K_1-2c,~~~~\mu_1=\mu+1.
\end{equation}
Thus, the energy eigenvalues can be expressed in terms of the parameters given above
\begin{equation}\label{72}
  E_{n}=\pm \frac{1}{a}\sqrt{(n+\mu+1/2)^{2}-(\mu+1/2)^{2}}
\end{equation}
where one can say that $j=n+\mu$  to compare our results with those found in \cite{yadav}.

\section{Conclusions}
Our findings in this paper point to the fact that the exact solutions for a given system which is relativistic can be obtained using the similar techniques used
in non-relativistic quantum  mechanics. Especially, considering the massless particle dynamics, the latest trends in relativisitic quantum mechanics can bring new bound state problems which are not solved yet such as the Dirac equation in a curved space-time which has a toroidal geometry. Because the metric
contains a more general trigonometric function which is $R(x)=c+a\cos x$, the Klein-Gordon-like equations obtained from the couple of first order Dirac equations are not  familiar  which are generally known in relativistic quantum mechanics. In this problem, the Fermi velocity is chosen as a non-constant function which is expressed in terms of the point transformation function in our solutions, after then, solvable potentials are derived using the superpotential suggestions. In the equal inner and outer radius case, one of the partner potentials is trigonometric P\"{o}schl-Teller potential while the other one is including the rational terms. We have obtained the solutions of the partner potentials for each system (\ref{8}) and (\ref{9}). For the different radius values of the torus surface, as a more general case, one of the partner potential is found as not solvable rational function
which includes beta function  while the other one is trigonometric P\"{o}schl-Teller potential. In the next case, unsolvable partner potential  is obtained in tems of the Appel hypergeometric functions. We also note that these unsolvable potentials share the same energy with the trigonometric P\"{o}schl-Teller potential. In the final section of this work, operators of the Lie algebra $iso(2,1)$
are found in order to express the Casimir operator with the potential functions like the extended trigonometric P\"{o}schl-Teller potentials which are
(\ref{20}) and (\ref{21}) given in  the $a=c$ case. Finally we note that the Dirac equation on the toroidal spacetime problem can lead to obtain
 more general potential families.

\end{document}